\documentclass[a4paper]{article}

\usepackage{INTERSPEECH2020}
\usepackage{multirow}
\usepackage{url}
\usepackage{hyperref}
\usepackage{commath}
\usepackage{array}
\usepackage{array,booktabs,caption,ragged2e}
\usepackage{tabularx}
\usepackage{pbox}

\newcolumntype{P}[1]{>{\centering\arraybackslash}p{#1}}
\newcolumntype{M}[1]{>{\centering\arraybackslash}m{#1}}

\title{A Further Study of Unsupervised Pre-training for Transformer Based Speech Recognition}
\name{Dongwei Jiang, Wubo Li, Ruixiong Zhang, Miao Cao, Ne Luo, Yang Han, Wei Zou, Xiangang Li}
\address{
  AI Labs, Didi Chuxing, Beijing, China}
\email {\{jiangdongwei, liwubo, zhangruixiong, caomiao, luone, hanyang, zouwei, lixiangang\}@didiglobal.com}

\begin{document}

\maketitle
\begin{abstract}
	Building a good speech recognition system usually requires large amounts of transcribed data, which is expensive to collect. To tackle this problem, many unsupervised pre-training methods have been proposed. Among these methods, Masked Predictive Coding achieved significant improvements on various speech recognition datasets with BERT-like Masked Reconstruction loss and Transformer backbone. However, many aspects of MPC have not been fully investigated. In this paper, we conduct a further study on MPC and focus on three important aspects: the effect of pre-training data speaking style, its extension on streaming model, and how to better transfer learned knowledge from pre-training stage to downstream tasks. Experiments reveled that pre-training data with matching speaking style is more useful on downstream recognition tasks. A unified training objective with APC and MPC provided 8.46\% relative error reduction on streaming model trained on HKUST. Also, the combination of target data adaption and layer-wise discriminative training helped the knowledge transfer of MPC, which achieved 3.99\% relative error reduction on AISHELL over a strong baseline. 
		
\end{abstract}


\section{Introduction}
\label{sec:intro}
	Current industrial end-to-end automatic speech recognition (ASR) systems rely heavily on large amounts of high quality transcribed audio data. However, transcribed data take substantial effort to obtain in industrial applications, while at the same time, a lot of un-transcribed data exist in online systems and cost little to collect. It is worthwhile to explore how to effectively use un-transcribed data to improve the performance of speech recognition systems when labeled data are limited. 

	Recently, unsupervised pre-training has shown promising results in several areas, including Computer Vision (CV) \cite{DBLP:conf/iccv/DoerschGE15},
	 Natural Language Processing (NLP) \cite{DBLP:conf/naacl/DevlinCLT19} 
	 and so on. One work that stands out among these methods is Bidirectional Encoder Representations from Transformers (BERT) \cite{DBLP:conf/naacl/DevlinCLT19}, which used a Masked Language Model (MLM) pre-training objective and obtained new state-of-the-art results on eleven NLP benchmarks.	 

	In speech area, researchers also proposed many unsupervised pre-training algorithms. Contrastive Predictive Coding (CPC) \cite{oord2018representation} extracts representation from data using contrastive loss that requires distinguishing a true future audio sample from negatives. There are multiple concurrent approaches that generalize this approach \cite{baevski2019vqwav2vec, Baevski2019EffectivenessOS} and applied it in learning speaker representation \cite{Ravanelli_2019}, extracting speech representation \cite{Schneider_2019, Pascual_2019}, performing various speech-related tasks like speech recognition \cite{Schneider_2019, Kawakami2020LearningRA, Rivire2020UnsupervisedPT}, speech emotion recognition \cite{lian2018improving} and so on. \cite{Chung_2019, Chung2019GenerativePF, chung2020improved} proposed Autoregressive Predictive Coding (APC) objective that predicts unseen future frame given past frames and also achieved good results on phonetic classification, speech recognition, and speech translation. Some other work \cite{Liu2019MockingjayUS, Wang2020UnsupervisedPO, Song2019SpeechXLNetUA, Jiang2019ImprovingTS} got motivation from NLP and applied similar methods on speech tasks.

	Among these methods, Masked Predictive Coding (MPC) \cite{Jiang2019ImprovingTS} achieved significant improvements on state-of-the-arts Transformer based speech recognition models on various datasets without introducing any additional parameters for the speech recognition model. However, many aspects of MPC have not been fully explored:
	 
	\begin{itemize}
		\item Speaking style has a strong impact on performance of ASR systems \cite{benzeghiba2007automatic, weintraub1996effect}. Whether speaking style of pre-training data would affect the performance of downstream tasks has not been fully investigated.
		\item The ability to perform streaming recognition is important for ASR systems. MPC may not work for streaming recognition because it requires bidirectional context for predictive coding. How MPC can contribute to streaming models is worth exploring.
		\item Despite abundant work \cite{Howard2018UniversalLM, Chronopoulou_2019, Sun2019HowTF} on NLP about knowledge transfer between pre-trained model and downstream task, there are few work exploring how to perform better knowledge transfer in the area of speech.
	\end{itemize}
  In this paper, we investigate these aspects of MPC and discuss how we can extend MPC for better speech recognition.

\section{Masked Predictive Coding}

\begin{figure}[]
        \centering
	\centerline{\includegraphics[clip, trim=2cm 2.5cm 1cm 2cm, scale=0.41]{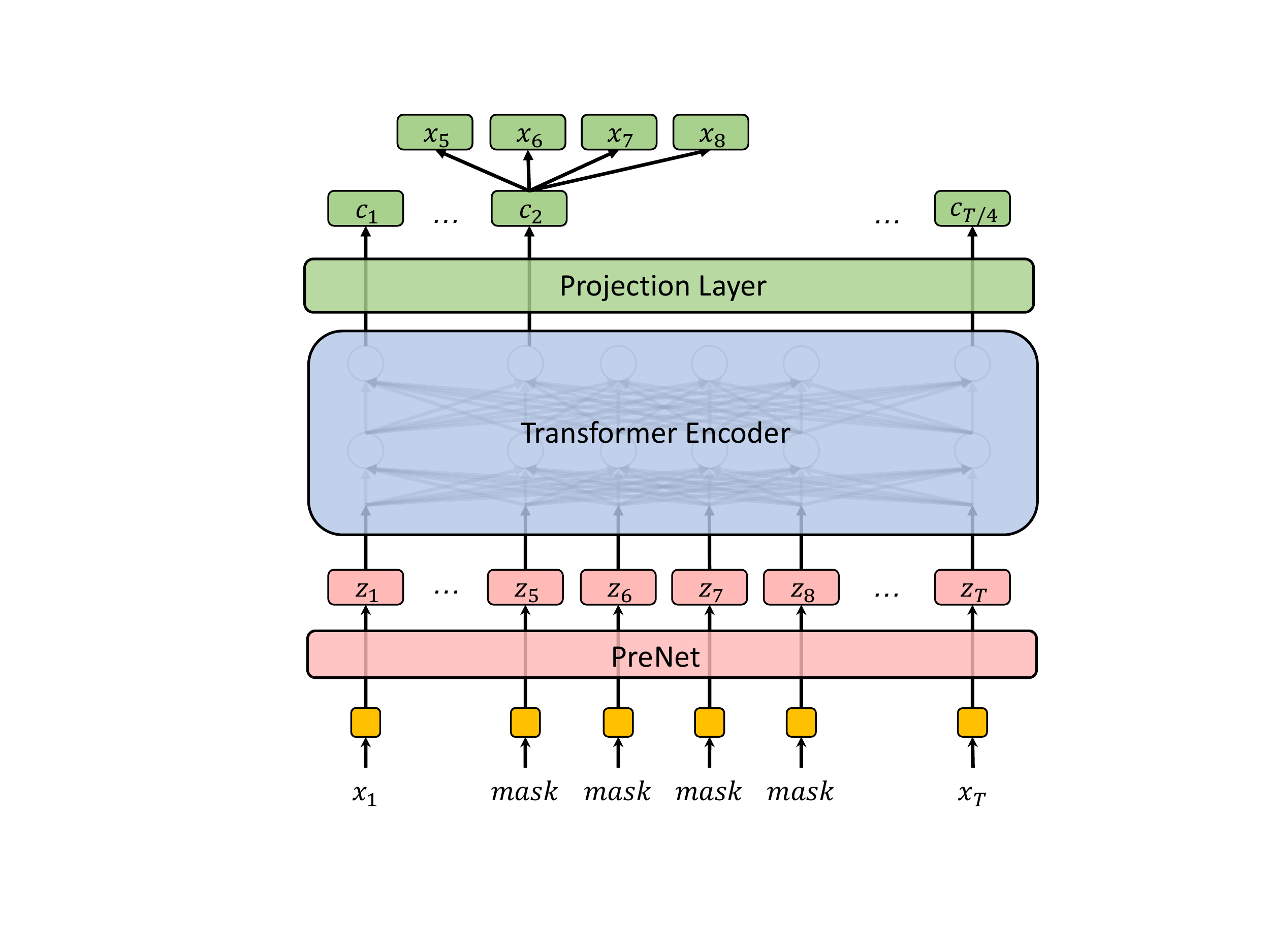}}
	\vspace{-0.3cm}
	\centerline{}\medskip
	  \vspace{-0.5\baselineskip}

  	\caption{Masked Predictive Coding with four-fold downsample}
  	\label{masked_pc}
	\end{figure}

	Masked Predictive Coding (MPC) \cite{Jiang2019ImprovingTS} uses Masked Reconstruction objective to perform predictive coding on Transformer based models. As depicted in Fig. \ref{masked_pc}, the model structure of MPC is essentially the encoder part of Transformer based speech recognition model plus a single fully-connected projection layer. During training, masks are applied on input FBANK features before feeding into the encoder. The training objective is L1 loss computed between masked input FBANK features and projected encoder output at corresponding position. The local smoothness of speech makes the task of predicting adjacent frames too easy. So, in the current setup of MPC, we divide input features into chunks of four frames and apply mask on chunks with a probability of 15\%. Unlike BERT, MPC adopted dynamic masking proposed in \cite{liu2019roberta} where the masking pattern is generated every time a sequence is fed to the model. 
	
	One unique characteristic of sequence-to-sequence with attention ASR models is it usually applies downsampling in the encoder. Previous research showed temporal pooling encourages effective encoding in different temporal resolution and makes alignments in the decoding easier \cite{ChanJLV16}. When temporal downsampling is applied, with input feature $X$ $\in$ $\mathbb{R}^{t \times d}$ and temporal downsampling rate $r$, Transformer encoder projects the output of last encoder layer to dimension $X_{0}$ $\in$ $\mathbb{R}^{t / r, d \times r}$. We then reshapes it back to same shape as input feature for MPC loss computation.

The overall training procedure of MPC consists of two stages, pre-training on unsupervised data and fine-tuning on supervised data. In the pre-training stage, MPC performs predictive coding directly on FBANK input and encoder output. After the unsupervised pre-training procedure, we remove the additional projection layer for predictive coding and add Transformer decoder for fine-tuning on downstream ASR tasks. All model parameters are end-to-end trainable in the fine-tuning stage. The work perhaps most similar to ours is Mockingjay \cite{Liu2019MockingjayUS}, which also employed Transformer encoder with Masked Reconstruction loss. But their work mainly used pre-trained model as a feature extractor while MPC works more like BERT and focuses on obtaining a good parameter initialization. Also, unlike \cite{Liu2019MockingjayUS} and other previous work on predictive coding \cite{Schneider_2019, Chung_2019}, our setup does not introduce any additional parameters into speech recognition model.


\section{Methods}
\label{sec:proposed_method}

	\subsection{MPC for streaming models}
		To apply MPC in streaming models, the Transformer encoder needs to be restricted to only use information that has appeared before. Though some previous work \cite{Sainath2018ImprovingTP, Miao2020TransformerbasedOC} employed chunk-wise splitting for streaming models, in this paper, we simply changed self-attention mask on Transformer encoder to make the whole model stream-able. Specifically, we use a triangular matrix for self-attention mask \textbf{M} in encoder, where the upper triangular part is set to ${-\infty}$, and the other elements to 0.

		Recent work on APC \cite{Chung2019GenerativePF} got impressive results on downstream tasks with Transformer decoder backbone. 
		Inspired by \cite{Dong2019UnifiedLM}, we also propose to use a unified training objective that combines MPC and APC. 
		During training, with probability \textbf{p}, we apply triangular matrix on Transformer encoder and use APC objective, with probability \textbf{1 - p}, we use the Transformer encoder as-is with MPC objective. 
		 This parameter sharing framework has the advantage of making the learned speech representations more general because they are jointly optimized for different pre-training objectives where context is utilized in different ways.

	\subsection{Knowledge transfer for MPC}
	  For speech recognition task in a specific domain, its data distribution may be a lot different from data used for MPC pre-training. Directly using MPC model in fine-tuning stage might cause degradation in performance, even catastrophic forgetting \cite{goodfellow2013empirical, Kirkpatrick_2017}. To deal with this problem, we followed previous work \cite{Howard2018UniversalLM, Sun2019HowTF} and adopted \textbf{target data adaption} to fine-tune MPC model on data of target task before the fine-tuning stage.
		
	  It is well-known that different layers of neural network capture different types of information, and the transferability of different layers also varies a lot \cite{Yosinski2014HowTA, Liu2019LinguisticKA}.
	  Previous work \cite{Howard2018UniversalLM} made use of these findings by assigning lower learning rates to layers that are more general for downstream tasks and achieved promising results.
	  To adapt their findings to MPC, we first used probing task \cite{Shi2016DoesSN, Adi2017FinegrainedAO} on pre-trained model to find out which layers of Transformer encoder are more useful downstream speech recognition tasks. After that, \textbf{layer-wise discriminative training} \cite{Howard2018UniversalLM} is used to assign different learning rates to each layer of Transformer encoder and adapt them to different extents for better knowledge transfer.

	  Transfer Learning with single-step auxiliary loss \cite{Chronopoulou_2019} is yet another transfer learning approach that adopted multi-task learning perspective via the addition of pre-training objective in the fine-tuning stage. With the same idea, we also added MPC loss in the fine-tuning stage. This way, the joint loss in the fine-tuning stage became the weighted sum of task-specific loss $L_{Attn}$, $L_{CTC}$ and auxiliary MPC loss $L_{MPC}$:
		\begin{equation}
			L = \alpha_{attn} * L_{Attn} + \beta_{ctc} * L_{CTC} + \gamma_{mpc} * L_{MPC},
		\end{equation}
	  For convenience, we name this method \textbf{multi-task with MPC} through out this paper. Masked input and output for MPC are also added during training.

\section{Experiments}
\label{sec:experiments}

	\subsection{Data}
	\label{sec:Pre-training data}
	For Mandarin pre-training, we first collected reading type dataset OpenMandarin as described in \cite{Jiang2019ImprovingTS}. Note unlike \cite{Jiang2019ImprovingTS}, we did not include HKUST and AISHELL in OpenMandarin dataset in this work because we found self pre-trained MPC also improves performance. OpenMandarin contains about 1150 hours of speech in total. 
	 To understand the impact of pre-training data speaking style, our internal reading type dataset Didi Dictation and spontaneous type dataset Didi Callcenter were also included. Didi Dictation is collected from our internal mobile dictation application while Didi Callcenter is collected from phone calls between our user and customer service staff. We randomly selected 5000 hours of Didi Callcenter and 5000 hours of Didi Dictation for MPC pre-training. For fair comparison with Open Mandarin, we also added another pre-training dataset by randomly selecting 1150 hours of Didi Callcenter and name it Didi Callcenter - 1K.
	 For English pre-training data, reading type dataset Librispeech \cite{Panayotov2015LibrispeechAA} and spontaneous type dataset Fisher \cite{Cieri2004TheFC} are used.	To provide a fair comparison on data size, we created a new pre-training dataset by randomly selecting only 960 hours of Fisher dataset and name it Fisher - 1K. Detailed information of pre-training corpora is provided in Table \ref{datasets}.

	Fine-tuning experiments were conducted on HKUST, AISHELL-1 and Switchboard. The speaking style of HKUST and Switchboard is spontaneous while the speaking style of AISHELL is reading. Speed perturbation of 0.9 and 1.1 was used on the training data. Mandarin characters are used as modeling units for HKUST and AISHELL like described in \cite{ZouJZYL18}, while 2000 BPE subwords \cite{Zenkel2018SubwordAC} is used for experiments on Switchboard. 
		
	The sample rate of HKUST and Switchboard is 8000 Hz, which is lower than some pre-training data. To alleviate the possible influence, we kept the sample rate of target dataset unchanged and downsampled pre-training data with higher sample rate to 8000 Hz when needed.

	\subsection{Experimental setups}
	\label{sec:model_structure}
	
		For Transformer based full-sequence models, we followed model structure of previous work \cite{karita} with $e=12$, $d=6$, $d_{model}=256$, $d_{ff}=1280$ and $d_{head}=4$. A source sequence X is first fed into a prenet consisting of two-layer CNN with 256 channels, stride size 2 and kernel size 3 and transformed to subsampled sequence $X_{0}$ $\in$ $\mathbb{R}^{n^{sub} \times d^{attn}}$ before feeding into Transformer.  For Transformer based streaming models, two representative work with CTC \cite{Salazar2019SelfattentionNF} and transducer loss \cite{Yeh2019TransformerTransducerES} are slightly modified and used in this work. We used the same prenet and encoder structure as Transformer based models. A two-layer Transformer decoder is used as prediction network for transducer.

		 In pre-training stage, all models are trained with a total batch size of 256 for about 100 epochs.
		 We used Adam optimizer and varied learning rate with warmup schedule \cite{VaswaniSPUJGKP17} according to the formula: 
 		\begin{equation}
	 		lrate = k * d_{model}^{0.5} * min(n^{-0.5}, n*warmup\_n^{-1.5}),
 		\end{equation}
		 where $n$ is the step number. $k = 0.5$ and $warmup\_n = 5000$ were chosen for all pre-training experiments. 
		 
		 In the fine-tuning stage, all models are trained with a total batch size of 128. The same warmup schedule is used except we changed $n$ to 1.0 and $warmup\_n$ to 25000. For Transformer based models, the Attention-CTC multi-task training objective is used, with weights determined on development set. 
		 Models for HKUST and AISHELL is trained for 50 epochs while model for Switchboard is trained for 100 epochs. Label smoothing of 0.1 and weight decay of $1e^{-5}$ are applied for all fine-tuning experiments. SpecAugument \cite{Park2019SpecAugmentAS} is applied when training for Switchboard.

		  In the decoding stage, we selected 10 models with lowest error rates on the validation set and averaged their parameters.	Beam search with Attention-CTC joint decoding and RNN language model \cite{Kim_2017} is used for Transformer based models, with weights determined using grid search on development set. The beam size is 10 for HKUST and AISHELL for Transformer based models and 20 for Switchboard. WFST with word-level language model is used for decoding with CTC.

			 We made our code publicly available for reproducibility at \nolinkurl{https://github.com/athena-team/athena/tree/mpc_improvement}.

			 \begin{table}[]
				\centering			
				\caption{Details of pre-training corpora}
				\begin{tabular}{l|l|l}
					\toprule
				Datasets		  		        &Hours	 						&Speaking Style 				\\ \midrule 
                Open Mandarin 					&1150 							&Reading 					\\ 
                Didi Callcenter - 1K 			&1150 							&Spontaneous						\\
				Didi Callcenter 				&5000 							&Spontaneous						\\
				Didi Dictation                  &5000 							&Reading						\\
                Librispeech					    &960					        &Reading						\\
                Fisher - 1K				        &960 							&Spontaneous						\\ \bottomrule
				\end{tabular}
				\vspace{-0.5\baselineskip}
				\label{datasets}
			\end{table}

	\subsection{Effect of pre-training data speaking style}
	\label{sec:pre-training for HKUST and AISHELL-1}

		The results on HKUST and AISHELL with different pre-training data are listed in Table \ref{loss_HKUST}. Our baseline result without MPC matches the strong baseline in \cite{karita}. Comparing relative error reduction of HKUST with same amounts of pre-training data, it is obvious MPC models pre-trained with matching speaking style data achieved lower error rates for downstream tasks. 

	  Open Mandarin and Didi Dictation are both used in pre-training for HKUST and AISHELL. From Table \ref{loss_HKUST}, we can also find the relative error reduction using the same pre-training data is bigger with matching fine-tuning data. Interestingly, the error reduction Didi Callcenter - 1K brings on HKUST is even bigger than Didi Dictation, which suggests speaking style has a bigger impact than pre-training data size in some cases.

	\begin{table}[]
		\centering
		\setlength{\tabcolsep}{5pt} 
		\caption{Character Error Rates(\%) and Relative Error Rates Reduction(\%) on HKUST and AISHELL test set}
		\begin{tabular}{l|l|l|l|llllll}
			\toprule
		Task                        &Pre-training Data        &Hours    &CER			  &RERR	\\ \midrule
		\multirow{6}{*}{HKUST}      &-				          &-        &23.3  		   	  & - 			    					\\
		                            &HKUST					  &170      &22.8 		   	  & 2.14			    		    \\
	                                &Open Mandarin (8k)	      &1150     &22.7 	 		  &	2.58		\\ 
		                            &Didi Callcenter - 1K     &1150	    &22.0 	 	      & 5.58			\\ 
		                            &Didi Callcenter  	      &5000     &\textbf{21.5}    & \textbf{7.73}			\\ 
		                            &Didi Dictation	(8k)	  &5000     &22.1 	 		  & 5.15			\\ \midrule
		\multirow{4}{*}{AISHELL}    &-				          &-        &6.82 		   	  & - 			    					\\
		                            &AISHELL			      &178      &6.61 		   	  & 3.07			    		    \\
		                            &Open Mandarin            &1150     &6.38 	 		  & 6.45			\\ 
		                            &Didi Dictation	    	  &5000     &\textbf{6.26} 	  & \textbf{8.21}			\\
		\bottomrule
		\end{tabular}
	\label{loss_HKUST}
	\end{table}


	\begin{table}[]
			\centering
			\caption{Word Error Rates(\%) and Relative Error Rates Reduction(\%) on Switchboard and CallHome test set}
			\begin{tabular}{l|l|ll|llllll}
				\toprule
			\multirow{2}{*}{Pre-training Data}   &\multirow{2}{*}{Hours} 	 &\multicolumn{2}{c|}{WER}      &\multicolumn{2}{c}{RERR}     \\
			                                     &                           &SWB		     &CH            &SWB		    &CH \\ \midrule			
			-					                 &-                          &8.8            &17.8 			&-              &-    					\\
			Switchboard		                     &260	                     &8.5 	         &17.4 			&3.41 	        &2.25    		    \\
			Fisher - 1K			                 &960                        &\textbf{8.0}   &\textbf{16.2} &\textbf{9.09}  &\textbf{8.99}    \\
			Librispeech (8k)	                 &960	                     &8.4	         &17.2			&4.55	        &3.37   		    \\
			\bottomrule

			\end{tabular}
			\vspace{-0.5\baselineskip}
			\label{loss_switchboard}
	\end{table}

		Experiments were also conducted on English ASR database Switchboard.  Our baseline results on Switchboard is also on par with previous work \cite{karita}.  As shown in Table \ref{loss_switchboard}, Switchboard got better results with spontaneous type pre-training data Fisher - 1K than with Librispeech, which further confirms our findings on speaking style.

	\subsection{MPC for streaming model}
	\label{sec:streaming models}
	  We first tested the effect of directly initializing streaming model with MPC. As shown in Table \ref{streaming_base}, pre-trained MPC brings consistent gains on streaming models for HKUST and AISHELL. MPC model pre-trained on more data obtained better results on streaming models, which is also in line with the conclusion for full-sequence models. Though not specifically optimized, MPC pre-training is also useful for streaming models, which suggests the parameter initialization obtained by MPC is helpful for both uni-directional and bidirectional models.

		\begin{table}[t!]
			\setlength{\tabcolsep}{5pt} 
			\centering
			\caption{Character Error Rates(\%) and Relative Error Rates Reduction(\%) for uni-directional CTC and uni-directional RNN-T with pre-trained MPC}
			\begin{tabular}{l|l|l|l|l}
				\toprule
	Task		            &Model				      & Pre-training Data	    & CER	           & RERR 	\\ \midrule
	\multirow{6}{*}{HKUST}  &\multirow{3}{*}{CTC}	  & -                       & 29.3	           & -       \\
						    &		                  & Didi Callcenter - 1K    & 28.0             & 4.56       \\ 
							&		                  & Didi Callcenter         & \textbf{27.6}    & \textbf{5.80}        \\ \cmidrule{2-5}
			                &\multirow{3}{*}{RNN-T}	  & -	                    & 28.1	           & -	      \\ 
							&		                  & Didi Callcenter - 1K    & 26.9             & 4.43       \\ 
							&		                  & Didi Callcenter         & \textbf{26.6}	   & \textbf{5.20}	      \\  \midrule
	\multirow{4}{*}{AISHELL}&\multirow{2}{*}{CTC}	  & -                       & 9.91 	           & -		    \\
							&		                  & Open Mandarin           & \textbf{9.42}	   & \textbf{4.91}       \\ \cmidrule{2-5}
			                &\multirow{2}{*}{RNN-T}	  & -	                    & 9.43 	           & -		    \\
							&		                  & Open Mandarin           & \textbf{8.88}    & \textbf{5.83}		\\ \bottomrule
			\end{tabular}
			\label{streaming_base}
		\end{table}

		To further improve performance of MPC on streaming models, we tried to combine APC with MPC in pre-training stage. Previous work on APC showed a future prediction step of 5 gave best results on Transformer decoder \cite{Chung2019GenerativePF}.
		We implemented APC with the same future prediction step and set switching probability \textbf{p} to 0.5. Experiments were conducted on HKUST dataset with Transformer CTC model and the results of different pre-training objectives are presented in Table \ref{different_prob}. 
		When used alone, APC and MPC got similar improvements. Combining APC with MPC results in consistent gains over them, which echoes with the findings in \cite{Dong2019UnifiedLM}. 

		\begin{table}[]
			\centering
			\caption{MPC + APC is the model pre-trained with APC 50\% of time and MPC 50\% of time. Relative Error Rates Reduction(\%) is calculated with HKUST baseline without MPC}
			\begin{tabular}{l|l|l|l}
				\toprule
			Pre-training Data		                          & Objective	      & CER	           & RERR	\\ \midrule
			\multirow{3}{*}{Didi Callcenter - 1K}             & MPC               & 28.0           & 4.56      \\  
			                                                  & APC               & 27.8	       & 5.12      \\
			                                                  & MPC + APC         & \textbf{27.2}  & \textbf{7.32}      \\ \midrule
		    \multirow{3}{*}{Didi Callcenter}                  & MPC               & 27.6           & 5.80	    \\ 
															  & APC               & 27.9           & 4.72	    \\ 
			                                                  & MPC + APC         & \textbf{26.8}  & \textbf{8.46}	    \\ \bottomrule
			\end{tabular}
			\vspace{-0.5\baselineskip}
			\label{different_prob}
		\end{table}

	\subsection{Knowledge transfer for MPC}
	\label{sec:knowledge transfer}

\begin{figure}[]
			\begin{minipage}[b]{0.9\linewidth}
				\centering
				\centerline{\includegraphics[clip, trim=0.5cm 7.5cm 2cm 8.5cm, scale=0.35]{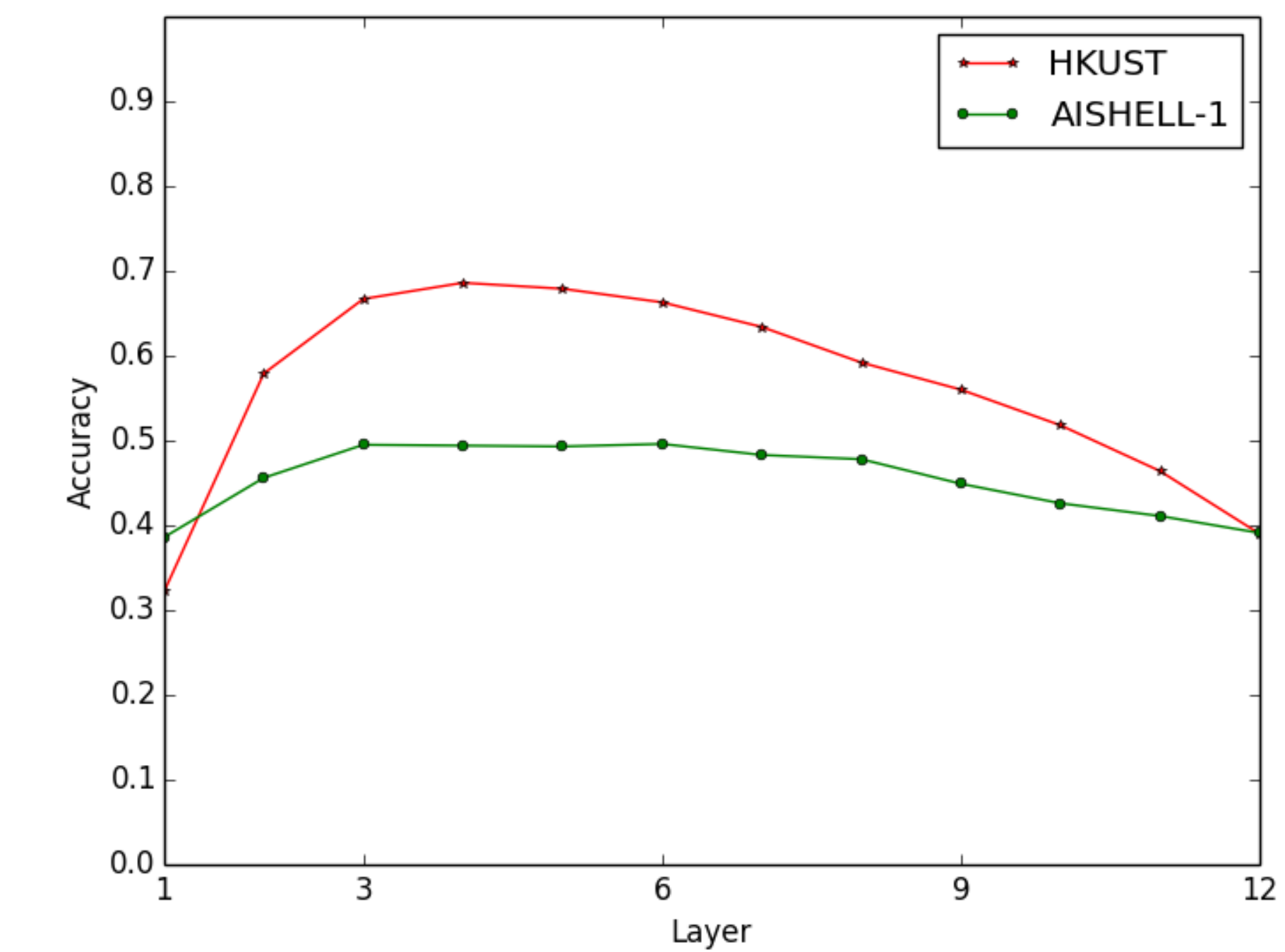}}
			\end{minipage}

			\caption{CER(\%) with each layer of pre-trained Transformer encoder. Results on HKUST is pre-trained with Didi Callcenter and results on AISHELL is pre-trained with Open Mandarin}
			\label{fig:probing_task}
		\end{figure}

		Each layer of pre-trained MPC encoder captures different features of the input speech. To get a quantitative measure of how important layer \textit{l} is, for each layer, we create a new Transformer with \textit{l} layers of encoder and 6 layers of decoder. The encoder for the new Transformer is frozen and initialized from pre-trained MPC model while the decoder is trainable. As shown in Fig \ref{fig:probing_task}, features from the middle layers of pre-trained Transformer encoder are generally more helpful than features from top and bottom layers. In the fine-tuning stage, we propose to set the learning rates of encoder layers discriminatively by multiplying the learning rate of layer \textit{l} with $ {\lambda}^{\norm{\textit{l} - {\theta}} } $, where $\lambda$ $\in$ (0, 1). To fit the accuracy curve above, we set $\lambda$ to be 0.95 and $\theta$ to be 5.5.
		This way, the parameters of middle layers are updated more slowly and knowledge from pre-trained MPC model is best retained. 
		Different $\lambda$ and $\theta$ values have also been tested and no significant difference is observed. Using this schedule on baseline systems without MPC doesn't lead to better accuracy either.

		The results for target data adaptation and layer-wise discriminative training are listed in Table \ref{knowledge_transfer_step}. Two baseline results we used are HKUST pre-trained with Didi Callcenter and AISHELL pre-trained with Open Mandarin. Using target data adaption or layer-wise discriminative training alone doesn't help the knowledge transfer of MPC very much. But when combined together, they provide consistent gains on downstream tasks.

		For multi-task with MPC, the MPC loss should contribute more to the joint loss to facilitate knowledge transfer in the first few epochs. As training proceeds, the task-specific component of the loss function becomes more important and $\gamma_{mpc}$ should be decreased. In this work, we empirically found it work well to set initial $\gamma_{mpc}$ to 0.2 and decrease it by half every 5 epochs. The result in Table \ref{knowledge_transfer_step} shows multi-task with MPC gets slight improvements over baseline.
		
		\begin{table}[]
			\centering
			\caption{Results on HKUST and AISHELL with different knowledge transfer methods. HKUST is pre-trained with Didi Callcenter. AISHELL is pre-trained with Didi Dictation. Target Data + Layer-wise is the combination of target data adaption and layer-wise discriminative training}
			\begin{tabular}{l|l|l|llll}
			\toprule
			Task	                   &Knowledge Transfer Methods  &CER 	       &RERR               \\ \midrule
            \multirow{5}{*}{HKUST}     &-                           &21.5	       & -              \\
									   &Target Data Adaption        &21.2          & 1.40               \\
									   &Layer-wise Discriminative   &21.3          & 0.93               \\
									   &Target Data + Layer-wise    &\textbf{20.8} &\textbf{3.26}               \\
									   &Multi-Task MPC              &21.1          & 1.86               \\ \midrule
			\multirow{5}{*}{AISHELL}   &-                           &6.26	       & -              \\
									   &Target Data Adaption        &6.23          & 0.48               \\
								       &Layer-wise Discriminative   &6.21          & 0.80               \\
									   &Target Data + Layer-wise    &\textbf{6.01} &\textbf{3.99}               \\
									   &Multi-Task MPC              &6.13          & 2.08               \\ \bottomrule
			\end{tabular}
			\vspace{-0.5\baselineskip}
			\label{knowledge_transfer_step}
		\end{table}

\section{Conclusion}
\label{sec:conclusion}
	In this work, we investigated three important aspects of MPC. Pre-training data with matching speaking style was found to be more useful on downstream recognition tasks. Using MPC directly on streaming models helps, but combining MPC with APC brings further improvements on streaming models. Also, the combination of target data adaption and layer-wise discriminative training provides consistent gains on knowledge transfer to downstream tasks. 


\bibliographystyle{IEEEtran}

\bibliography{mybib}

\end{document}